\documentclass[usenatbib]{mn2e}
\usepackage{graphicx,times}
\usepackage{natbib}
\usepackage{amssymb}
\usepackage{amsmath,bm}
\newcommand{\be}{\begin{equation}}
\newcommand{\ee}{\end{equation}}
\newcommand{\ba}{\begin{eqnarray}}
\newcommand{\ea}{\end{eqnarray}}

\newcommand{\bi}{\begin{itemize}}
\newcommand{\ei}{\end{itemize}}
\newcommand{\bfi}{\begin{figure}
\epsfxsize=9cm
\epsffile}
\newcommand{\efi}{\end{figure}}

\newcommand{\mnras}{MNRAS}
\newcommand{\apj}{ApJ}
\newcommand{\apjl}{ApJ}

\newcommand{\physrep}{Physics Reports}

\newcommand{\nat}{Nature}
\title[Lensing reconstruction through magnification]{Weak lensing
  reconstruction through cosmic magnification  I: a
  minimal variance map reconstruction}
\author[Yang and Zhang]{Xinjuan Yang$^{1,2}$\thanks{E-mail:yxj@bao.ac.cn} and
Pengjie
Zhang$^{3}$\\
$^{1}$National Astronomical Observatories,Chinese Academy of
Sciences,Beijing 100012,China. \\
$^{2}$Graduate University of Chinese Academy of Sciences, 19A,
Yuquan Road, Beijing, 100049, China. \\
$^{3}$Key Laboratory for Research in Galaxies and Cosmology,
Shanghai Astronomical Observatory, Nandan Road 80, Shanghai, 200030,
China}

\begin{document}
\maketitle
\begin{abstract}
We present a concept study on weak lensing map reconstruction
through the cosmic magnification effect in galaxy number density
distribution. We propose a minimal variance linear estimator to
minimize both the dominant systematical and statistical errors in
the map reconstruction. It utilizes the distinctively different flux
dependences to separate the cosmic magnification signal from the
overwhelming galaxy intrinsic clustering noise. It also minimizes
the shot noise error by an optimal weighting scheme on the galaxy
number density in each flux bin.  Our method is in principle
applicable to all galaxy surveys with reasonable redshift
information.  We demonstrate its applicability against the planned
Square Kilometer Array survey, under simplified conditions.  Weak
lensing maps reconstructed through our method are complementary to
that from cosmic shear and CMB and 21cm lensing. They are useful for
cross checking over systematical errors in weak lensing
reconstruction and for improving cosmological constraints.

\end{abstract}
\section{introduction}
Weak gravitational lensing has been established as one of the most powerful
probes of  the dark universe \citep{Refregier2003,Albrecht2006,Munshi08,Hoekstra2008}. It is rich in
physics and contains tremendous information on dark matter,  dark energy and
the nature of gravity at cosmological scales. Its modeling is relatively clean. At
 the  multipole $\ell< 2000$,
gravity is the dominant  force shaping the weak lensing power spectrum while
complicated gas physics only affects the lensing power spectrum at less than $
1\%$ level \citep{White04,Zhan04,Jing06,Rudd08}. This  makes the weak lensing
precision modeling feasible, through high
precision simulations.

Precision  weak lensing measurement is also promising.  The most
sophisticated and successful method so far  is to measure the cosmic
shear,  lensing  induced galaxy shape distortion.   After the first
detections in the year 2000
\citep{Bacon2000,Kaiser00,Waerbeke2000,Wittman00}, data  quality has
been  improved dramatically (e.g. \citealt{Fu2008}). Ongoing and
planed surveys, such as DES
\footnote{http://www.darkenergysurvey.org}, LSST
\footnote{http://www.lsst.org}, JDEM
\footnote{http://jdem.gsfc.nasa.gov}, and Pan-STARRS
\footnote{http://pan-starrs.ifa.hawaii.edu/public/science}, have
great promise for further significant improvement.

However, weak lensing reconstruction through cosmic shear  still
suffers from practical difficulties associated with galaxy shape.
These include shape measurement errors (additive and multiplicative)
\citep{Heymans2006, Massey2007} and the galaxy intrinsic alignment
\citep{Croft2000, Heavens2000, Jing2002, Hirata2004, Mandelbaum2006,
Hirata2007, Okumura2009a, Okumura2009b}.

An alternative method for weak lensing reconstruction is through
cosmic magnification, the lensing induced fluctuation in galaxy (or
quasar and any other celestial objects) number density
(\citealt{Menard2002} and references therein). Since it does not
involve galaxy shape, it automatically avoids all problems
associated with galaxy shape.

However, the amplitude  of cosmic magnification in galaxy number
density fluctuation is usually one or more orders of magnitude
overwhelmed by the intrinsic galaxy  number fluctuation  associated
with the large scale structure of the universe. Existing cosmic
magnification measurements
\citep{Scranton2005,Hildebrandt2009,Menard2010,Waerbeke2010}
circumvent this problem by cross-correlating two galaxy (quasar)
samples widely separated in redshift.  Unfortunately, the measured
galaxy-galaxy cross correlation is often dominated by the foreground
galaxy density-background cosmic magnification correlation and is
hence proportional to a unknown galaxy bias of foreground galaxies.
This severely limits its cosmology application.

The intrinsic galaxy clustering and cosmic magnification have
different redshift and flux dependence, which can be utilized to
extract cosmic magnification. \citet{Zhang2006} demonstrated that,
by choosing (foreground and background) galaxy samples sufficiently
bright and sufficiently far away, the measured cross correlation
signal can be dominated by the cosmic magnification-cosmic
magnification correlation, which is free of the unknown galaxy bias.
However, even for those galaxy samples, the foreground galaxy
density-background cosmic magnification correlation is still
non-negligible \citep{Zhang2006}. This again  limits its cosmology
application due to the galaxy bias problem.

\citet{Zhang2005} further showed that, since the galaxy intrinsic
clustering and cosmic magnification have distinctive dependence on
galaxy flux, cosmic magnification can be extracted by appropriate
weighting over the observed galaxy number density in each flux bins.
Weak lensing reconstructed from spectroscopic redshift surveys such
as the square kilometer array (SKA) in this way can achieve accuracy
comparable to that of cosmic shear of stage IV projects.  These
works demonstrate the great potential of cosmic magnification as a
tool of precision weak lensing reconstruction. Furthermore, since
cosmic shear and cosmic magnification are independent methods, they
would provide valuable cross-check of systematical errors in weak
lensing measurement and useful information on galaxy physics such as
galaxy intrinsic alignment.

\cite{Zhang2005} only discussed two limiting cases, completely deterministic
and completely stochastic biases with known flux dependence. In reality,
galaxy bias could be partly stochastic. Furthermore, the flux dependence of
galaxy bias is not given {\it a priori}. In this paper, we aim to investigate
a key question. Are we able to simultaneously measure both cosmic
magnification and the intrinsic galaxy clustering, given the galaxy number
density measurements in flux and  redshift bins?

The paper is organized as follows. In \S \ref{sec:estimator}, we
present the basics of cosmic magnification and derive a minimal
variance  estimator for weak lensing reconstruction through cosmic
magnification. In a companion paper we will discuss an alternative
method to measure the lensing power spectrum through cosmic
magnification. In \S \ref{sec:error}, we discuss various statistical
and  systematical errors associated with this reconstruction. In \S
\ref{sec:result}, we target at SKA to demonstrate the performance of
the proposed estimator. We discuss and summarize in \S
\ref{sec:conclusion}. We SKA specifications are specified in the
appendix \ref{sec:SKA}.   Throughout the paper, we adopt the WMAP
five year data \citep{Komatsu2009} with $\Omega_{\rm m}=0.26$,
$\Omega_{\rm \Lambda}=0.74$, $\Omega_{\rm b}=0.044$, $h=0.72$,
$n_{\rm s}=0.96$ and $\sigma_8=0.80$.

\section{A minimal variance linear estimator}
\label{sec:estimator}
 Weak lensing  changes the number density of background  objects, which
is called cosmic magnification. It involves two competing effects. A
magnification of solid angle of source objects, leading to dilution
of the source number density, and a magnification of the flux,
making objects brighter. Let $n(s,z_{\rm s})$ be the unlensed number
density at flux $s$ and redshift $z_{\rm s}$, and the corresponding
lensed quantity be $n^{\rm L}(s^{\rm L},z_{\rm s})$. Throughout the
paper the superscript ``L'' denotes the lensed quantity. The galaxy
number conservation then relates the two by
\begin{equation}
\label{eqn:n} n^{\rm L}(s^{\rm L},z_{\rm s})ds^{\rm
L}=\frac{1}{\mu}n(s,z_{\rm s})ds \ ,
\end{equation}
where $s^{\rm L}=s\mu$.  $\mu({\bm \theta},z_{\rm s})$ is the
lensing magnification at corresponding direction $\bm\theta$ and
redshift $z_{\rm s}$,
\begin{equation}
\mu({\bm \theta},z_{\rm s})=\frac{1}{\big(1-\kappa({\bm
\theta},z_{\rm s})\big)^2-\gamma^2({\bm \theta},z_{\rm
s})}\approx1+2\kappa({\bm \theta},z_{\rm s}) .
\end{equation}
The last expression has adopted the weak lensing approximation such
that the lensing convergence $\kappa$ and the lensing shear $\gamma$
are both much smaller than unity ($|\kappa|,|\gamma|\ll 1$).
$\kappa$ for a source at redshift $z_{\rm s}$ is related to the
matter overdensity $\delta_m$ along the line of sight by
\begin{equation}
\kappa({\bm{\theta}},z_{\rm s})=\frac{3H_0^2\Omega_{\rm
m}}{2c^2}\int_0^{\chi_{\rm s}}{\frac{D(\chi_{\rm s}-\chi)D(\chi)}
{D(\chi_{\rm s})}\delta_{\rm m}(z,{\bm{\theta}})(1+z)d\chi}\ .
\end{equation}
Here, $\chi$ and $\chi_{\rm s}$ are the radial comoving  distances
to the lens at redshift $z$ and the source at redshift $z_{\rm s}$,
respectively. $D(\chi)$ denotes the comoving angular diameter
distance, which is equaling to $\chi$ for a flat universe.

Taylor expanding Eq. \ref{eqn:n} around the flux $s^{\rm L}$, we
obtain
\begin{equation}
n^{\rm L}(s^{\rm L})=\frac{1}{\mu^2}n(\frac{s^{\rm L}}{\mu})\approx
n(s^{\rm L})[1+2(\alpha-1)\kappa+O(\kappa^2)] \ ,
\end{equation}
Where the parameter $\alpha$ is related to the logarithmic slope of the
luminosity function,
\begin{equation}
\alpha\equiv-\frac{s^{\rm L}}{n(s^{\rm L})}\frac{dn(s^{\rm
L})}{ds^{\rm L}}-1=-\frac{d\ln{n(s^{\rm L})}}{d\ln{s^{\rm L}}}-1 \ .
\label{a}
\end{equation}
Notice that $n(s^{\rm L})$ is related to the cumulative luminosity
function $N(>s)$ by $N(>s)=\int_{s}^\infty n(s^{\rm L})ds^{\rm L}$.
Weak lensing then modifies the galaxy number over-density to
\begin{equation} \label{delta}
\delta^{\rm L}_{\rm g}\simeq \delta_{\rm g}+2(\alpha-1)\kappa\equiv
\delta_{\rm g}+g\kappa\ .
\end{equation}
Here, $\delta_{\rm g}$ is the intrinsic (unlensed) galaxy number
over-density (galaxy intrinsic clustering). For brevity, we have
defined $g\equiv 2(\alpha-1)$ and will use this notation throughout
the paper. Obviously, to the first order of gravitational lensing,
the cosmic magnification effect can be totally described by the
cosmic convergence and the slope of galaxy number count.

The luminosity function averaged over sufficiently large sky area is
essentially unchanged by lensing, \be \langle n^{\rm
L}\rangle=\langle n\rangle(1+O(\langle
\kappa^2\rangle))\approx\langle n\rangle\ . \ee The last expression
is accurate to $0.1\%$ since $\langle \kappa^2\rangle\sim 10^{-3}$.
The above approximation is important in cosmic magnification. It
allows us to calculate $\alpha$ by replacing the unlensed (and hence
unknown) luminosity function $n$ with $n^{\rm L}$, the directly
observed one.  This means that $\alpha$ is also an observable and
its flux dependence is directly given by observations. We will see
that this is the key to extract cosmic magnification from galaxy
number density distribution.

\begin{figure}
\includegraphics[angle=270,width=84mm]{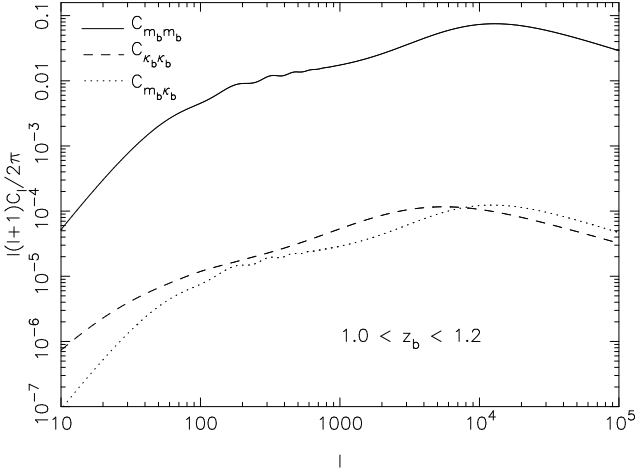}
\caption{Contaminations of the galaxy intrinsic clustering to the weak lensing
  reconstruction through cosmic magnification. The dominant contamination
  comes from the galaxy auto correlation, which is often several orders of magnitude
  larger than the lensing correlation. The galaxy intrinsic clustering
  also induces a galaxy-lensing cross correlation. This contamination can be
  comparable to the lensing signal. For a redshift bin $1.0<z_{\rm b}<1.2$, we
  plot the resulting matter power spectrum $C_{\rm m_bm_b}$, matter-lensing cross power
  spectrum $C_{\rm m_b\kappa_b}$ and the lensing power spectrum $C_{\rm \kappa_b\kappa_b}$. For
  galaxies with bias $b_{\rm g}\neq 1$, a factor $b_{\rm g}^2$ shall be applied to $C_{\rm m_bm_b}$
  and a factor $b_{\rm g}$ shall be applied to $C_{\rm m_b\kappa_b}$.  \label{fig:Cbb}}
\end{figure}

\begin{figure}
\includegraphics[angle=270,width=84mm]{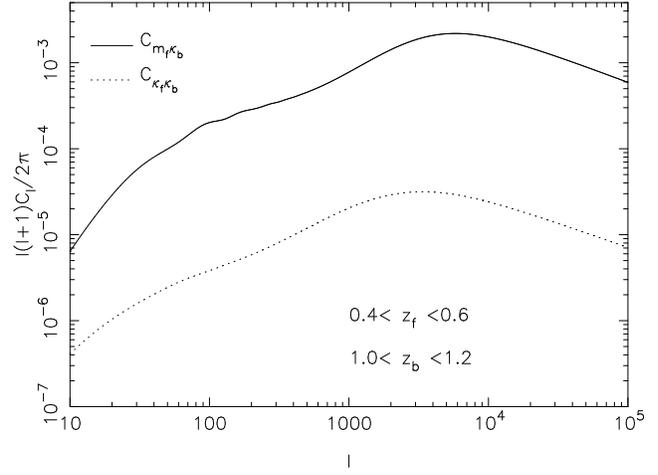}
\caption{Contamination of foreground galaxy intrinsic clustering to
the cross-correlation  weak lensing  power spectrum between
foreground and background redshift bins. The foreground galaxy
intrinsic clustering is related with the background cosmic
convergence and then induces a power spectrum $C_{\rm m_f\kappa_b}$.
For a couple of redshift bins $\left([0.4<z_{\rm f}<0.6,1.0<z_{\rm
b}<1.2]\right)$, we plot the weak lensing signal $C_{\rm
\kappa_f\kappa_b}$ and compare it with this main contamination
$C_{\rm m_f\kappa_b}$. The contamination overwhelms the signal by
one or more orders of magnitude.} \label{fig:Cbf}
\end{figure}

The biggest challenge in weak lensing reconstruction through cosmic
magnification is to remove the galaxy intrinsic clustering
$\delta_{\rm g}$. This kind of noise in cosmic magnification
measurement is analogous to the galaxy intrinsic alignment in cosmic
shear measurement.  But the situation here is much more severe,
since $\delta_{\rm g}$ is known to be strongly correlated over wide
range of angular scales and making it overwhelming the lensing
signal at virtually all angular scales (Fig. \ref{fig:Cbb} \&
\ref{fig:Cbf}).

To a good approximation, $\delta_{\rm g}=b_{\rm g}\delta_{\rm m}$,
where $\delta_{\rm m}$ is defined as matter surface over-density and
$b_{\rm g}$ is the galaxy bias. This is the limiting case of
deterministic bias. In general, for statistics no higher than second
order, $\delta_{\rm g}$ can be described with an extra parameter,
the galaxy-matter cross correlation coefficient $r$. $b_{\rm g}$ and
$r$ are defined through
\begin{eqnarray}
\label{eqn:br} b^2_{\rm g}(l,z)\equiv\frac{C_{\rm g}(l,z)}{C_{\rm
m}(l,z)}\ ,\ r(l,z)\equiv\frac{C_{\rm gm}(l,z)}{\sqrt{C_{\rm
g}(l,z)C_{\rm m}(l,z)}} \ .
\end{eqnarray}
Here, $C_{\rm g}$, $C_{\rm m}$ and $C_{\rm gm}$ are the galaxy,
matter and galaxy-matter angular power spectrum, respectively.
 On large scales corresponding to $ l\la k_{\rm L}\chi$ ($k_{\rm L}\sim\rm 0.1h/Mpc$),
the deterministic bias ($r=1$) is expected. However on nonlinear
scales, galaxy bias is known to be stochastic \citep{Pen1998,
  Dekel1999,Pen2003,Pen2004, Hoekstra2002, Fan2003}.

\subsection{The estimator}
The data we have are measurements of $\delta_{\rm g}^{\rm L}({\bm
\theta})$ at each angular pixel of each redshift and flux bin.
Throughout the paper we use the subscript ``$i$'' and ``$j$'' to
denote the flux bins. For convenience, we will work in Fourier
space. Then for a given redshift bin and a given multipole
$\bm{\ell}$, we have the Fourier transform of  the galaxy
over-density of the $i$-th flux bin, $\delta_{{\rm g},i}^{\rm
L}({\bm \ell})$. For brevity, we simply denote it as
$\delta_{i}^{\rm L}$ hereafter.  The $g$ factor of the $i$-th flux
bin is denoted as $g_{i}$. As explained earlier, it is a measurable
quantity. $b_{i}$ is the galaxy bias of the $i$-th flux bin.

We want to find an unbiased linear estimator of the form \be
\hat{\kappa}=\sum_{i} w_{i}\delta_{i}^{\rm L}\ , \ee Such that the
expectation value of $\hat{\kappa}$ is equal to the true $\kappa$ of
this pixel. Thus the weighting function $w$ must satisfy the
following conditions, \be \label{eqn:wg} \sum_{i} w_{i}g_{i}=1\ ,
\ee \be \label{eqn:wb} \sum_{i} w_{i}b_{i}=0\ . \ee Since the
measured $\delta_{i}^{\rm L}$ is contaminated by shot noise
$\delta_{i}^{\rm shot}$, we shall minimize the shot noise. This
corresponds to minimize \be \label{eqn:shot}
\left\langle\left|\sum_{i} w_{i}\delta^{\rm L}_{
i}-\kappa\right|^2\right\rangle=\left\langle\left|\sum_{i}
w_{i}\delta_{i}^{\rm shot}\right|^2\right\rangle=\sum_{i}
\frac{w_{i}^2}{\bar{n}_{i}}\ . \ee Here, $\bar{n}_{i}$ is the
average galaxy surface number density of the $i$-th flux bin, a
directly measurable quantity.

The three sets of requirement (Eq. \ref{eqn:wg}, \ref{eqn:wb} \&
\ref{eqn:shot}) uniquely fix the solution. Using the Lagrangian
multiplier method, we find the solution to be \be \label{eqn:w}
w_{i}=\frac{\bar{n}_{i}}{2}(\lambda_1g_{i}+\lambda_2b_{i})\ . \ee
Here, the two Lagrangian multipliers $\lambda_{1,2}$ are given by
\ba \label{eqn:lambda} \lambda_1&=&-\frac{2\sum
\bar{n}_{i}b_{i}^2}{(\sum \bar{n}_{i}b_{i}g_{i})^2-\sum
\bar{n}_{i}b_{i}^2\sum \bar{n}_{i}g_{i}^2} \ \nonumber,\\
 \lambda_2&=&\frac{2\sum \bar{n}_{i}b_{i}g_{i}}{(\sum
\bar{n}_{i}b_{i}g_{i})^2-\sum \bar{n}_{i}b_{i}^2\sum
\bar{n}_{i}g_{i}^2}\ . \ea  $w_{i}$ is invariant under a flux
independent scaling in $b_{i}$. For this reason, we only need to
figure out the relative flux dependence in the galaxy bias, instead
of its absolute value.

Despite the neat mathematical solution above, in reality we do not
know the galaxy bias {\it a priori}. We adopt a recursive procedure
to simultaneously solve $b_{i}$, $w_{i}$ and hence $\hat{\kappa}$.
\bi
\item The first step. In general,  the power spectrum is dominated
by the galaxy intrinsic clustering, namely \be \label{eqn:bi}
\left\langle |\delta_{i}^{\rm L}|^2\right\rangle=b_{i}^2C_{\rm
m_bm_b}+g_{i}g_{i}C_{\rm \kappa_b\kappa_b}+2b_{i}g_{i}C_{\rm
m_b\kappa_b }\simeq b_{i}^2C_{\rm m_bm_b}\ . \ee Here, $C_{\rm
m_bm_b}$ is the matter density angular power spectrum, $C_{\rm
\kappa_b\kappa_b}$ is the lensing power spectrum and $C_{\rm
m_b\kappa_b}$ is the cross-power spectrum, for the background
redshift bin. Hence a natural initial guess for $b_{i}$ is given by
the following equation, \be \label{eqn:b1} (b_{i}^{(1)})^2C_{\rm
m_bm_b}=\left\langle |\delta_{i}^{\rm L}|^2\right\rangle \ . \ee
Plug $b_{i}^{(1)}$ into in Eq. \ref{eqn:lambda} and Eq. \ref{eqn:w},
we obtain the weighting $w_{i}^{(1)}$, and then the first guess of
lensing convergence, $\kappa^{(1)}=\sum_{i}
w_{i}^{(1)}\delta_{i}^{\rm L}$.

$C_{\rm m_bm_b}$ is cosmology dependent, so one may think the bias
  reconstruction and hence the proposed lensing reconstruction are cosmology
  dependent. However, this is not the case.  As explained earlier, the
  weighting function $w_{i}$ does not depend on the   absolute value of
$b_{i}$.  So in the exercise, $C_{\rm m_bm_b}$ can be fixed to any
value of
  convenience in  determining the bias since it is independent with flux. In this sense, the bias reconstruction is
cosmology independent and is hence  free of uncertainties from
cosmological parameters.

\item The second step.  We subtract the lensing
contribution from the measurement $\delta^{\rm L}_{i}$, using
$\kappa^{(1)}$ constructed above. Our second guess for $b_{i}$ is
then \be b_{i}^{(2)}b_{i}^{(2)}C_{\rm
m_bm_b}=\left\langle\left|\delta_{i}^{\rm
L}-g_{i}\kappa^{(1)}\right|^2\right\rangle \ .
 \ee

 We then obtain $w^{(2)}_{i}$ and $\kappa^{(2)}$.  However,
this solution is still not exact. The expectation value of the r.h.s
is \be \left\langle \left|\delta_{i}^{\rm
L}-g_{i}\kappa^{(1)}\right|^2\right\rangle=\bigg(b_{i}-g_{i}\sum_{j}
w_{j}^{(1)}b_{j}\bigg)^2C_{\rm m_bm_b} \ . \ee So the expectation
value of $b_{i}^{(2)}$ is $b_{i}-g_{i}\sum_{j} w_{j}^{(1)}b_{j}$.
\item The iteration. We repeat the above step to obtain $b_{i}^{(p)}$ and $\kappa^{(p)}$
  ($p=3,\cdots$) until the iteration converges.  Since we know that the  lensing
contribution is sub-dominant and we start our iteration by
neglecting the lensing contribution, and since the flux dependence
of the intrinsic clustering and cosmic magnification ($b_{i}$ and
$g_{i}$) are different , such iteration should be stable and
converge. We numerically check it to be the case. \ei

The information of galaxy bias is not only encoded in the observed
power spectrum of the same flux bin, but also encoded in the cross
power spectra between different flux bins.  In our exercise, we do
not take this extra information into account.  So there are
possibilities for further improvement.

 In the above
description, we have neglected shot noise (or equivalently assumed
that it can be subtracted completely). In reality, we are only able
to subtract shot noise up to the limit of cosmic variance. Residual
shot noise introduces systematical error in the lensing
reconstruction, which we will quantify in next section.

Finally, we obtain the optimal estimator of cosmic convergence
\begin{equation}
\kappa^{(n)}=\sum_{i} w_{i}^{(n)}\delta_{i}^{\rm
L}=\kappa+\left(\sum_{i} w_{i}^{(n)}b_{i}\right)\delta_{\rm m} .
\label{eqn:k}
\end{equation}
Our estimator explicitly satisfies $\sum_{i}
w_{i}^{(n)}b^{(n)}_{i}=0$. However, since  $b_{i}^{(n)}$ can deviate
from its real value $b_{i}$, our estimator can be biased. This is
yet another source of systematical errors in our weak lensing
reconstruction through cosmic magnification. However, later we will
show that such systematical error is under control.

\subsection{The reconstructed weak lensing power spectrum}
From the reconstructed $\kappa$, we can reconstruct the lensing power
spectrum.   For the same redshift bin, we have
\begin{eqnarray}
&&C_{\rm bb}=\left\langle |\kappa^{(n)}_{\rm b}|^2\right\rangle
\\ \nonumber &&=C_{\rm \kappa_b\kappa_b}+\bigg(\sum_{i}
w_{i}^{\rm b}b_{i}^{\rm b}\bigg)^2C_{\rm m_bm_b}+2\bigg(\sum_{i}
w_{i}^{\rm b}b_{i}^{\rm b}\bigg)C_{\rm m_b\kappa_b}\ . \nonumber
\label{Cbb}
\end{eqnarray}
Throughout the paper we use the superscript or subscript ``b'' for
 quantities of background redshift bin and ``f'' for quantities of foreground
redshift bin, and we neglect the superscript ``(n)'' of weighting
$w_{i}^{(n)}$ from this section. For the cross correlation between
foreground and background populations, we have
\begin{eqnarray}
C_{\rm bf}&=&\left\langle\kappa^{(n)}_{\rm b}\kappa^{(n)\ast}_{\rm
f} \right\rangle=C_{\rm
  \kappa_f\kappa_b}+\left(\sum_{i} w_{i}^{\rm f}b_{i}^{\rm f}\right)C_{\rm
  m_f\kappa_b}\ .
\label{Cbf}
\end{eqnarray}
Here, $C_{\rm \kappa_f\kappa_b}$ is the lensing power spectrum
between foreground and background redshift bins, and $C_{\rm
m_f\kappa_b}$ is the cross power spectrum between foreground dark
matter and background lensing convergence. We have neglected the
correlation $C_{\rm m_fm_b}$ between foreground and background
matter distributions. It is natural for non-adjacent redshift bins
with separation $\Delta z\ge 0.1$, because of physical irrelevance.
For two adjacent redshift bins, there is indeed a non-vanishing
matter correlation. However, this correlation is also safely
neglected since  both the foreground and background intrinsic
clustering are sharply suppressed by factors $1/\sum_{i} w_{i}^{\rm
f,b}b_{i}^{\rm f,b}$, respectively.

\section{Statistical and systematical errors}
\label{sec:error} Our reconstruction method does not rely on priors
on galaxy bias, in this sense, it is robust. However, there is still
a number of statistical and systematical errors in the $\kappa$
reconstruction and weak lensing power spectrum reconstruction.  In
this section, we outline and quantify associated errors in the
lensing power spectrum.
\subsection{The statistical error}
\label{subsec:statisticalerror}
Our estimator minimizes the shot noise in the measurement. For auto
power spectrum $C_{\rm bb}$, the associated statistic measurement
error is
\begin{eqnarray}
\label{eqn:statistical} \Delta C_{\rm
bb}&=&\sqrt{\frac{1}{(2l+1)\Delta lf_{\rm
sky}}}\left\langle\left(\sum_{i} w_{i}^{\rm b}\delta_{i}^{\rm L}-\kappa_{\rm b}\right)^2\right\rangle \\
\nonumber &=&\sqrt{\frac{1}{(2l+1)\Delta lf_{\rm
sky}}}\sum_{i}\frac{(w_{i}^{\rm b})^2}{\bar{n}_{i}^{\rm b}}\ .
\end{eqnarray}
For the cross-power spectrum $C_{\rm bf}$, the statistical error is
\begin{eqnarray}
\Delta C_{\rm bf}&=&\sqrt{\frac{1}{(2l+1)\Delta lf_{\rm
sky}}}\sqrt{C^{\rm shot}_{\rm b}C^{\rm shot}_{\rm f}}
\\ \nonumber
&=&\sqrt{\frac{1}{(2l+1)\Delta lf_{\rm sky}}}\sqrt{\left(\sum_{i}
\frac{(w_{i}^{\rm b})^2}{\bar{n}_{i}^{\rm b}}\right)\left(\sum_{i}
\frac{(w_{i}^{\rm f})^2}{\bar{n}_{i}^{\rm f}}\right)}\ .
\end{eqnarray}
Throughout the paper we apply $\Delta l=0.2l$, and $f_{\rm sky}$ is
the sky coverage.

The above errors are purely statistical errors causing by shot noise
resulting from sparse galaxy distribution. For this reason, we call
them the weighted shot noise.  We reconstruct the $\kappa$ from the
measurements of lensed galaxy number over-density $\delta_{i}^{\rm
L}$.  We solve the galaxy bias and weighting function from the
observed galaxy power spectra $\langle|\delta_i^{\rm L}|^2\rangle$,
and this whole process is free of any given cosmological model. So
our proposed method to reconstruct the weak lensing map is the one
at the given sky coverage, with the right cosmic variance. Only when
we compare our measured lensing power spectra with their ensemble
average predicted by a given theory, we must add cosmic variance.
Since statistical errors arising from cosmic variance are
independent to shot noise, it can be taken into account
straightforwardly.

\subsection{Systematical errors}
\label{subsec:systematicalerror} We use the symbol $\delta C$ to
denote systematical errors in the lensing power spectrum
measurement.  We have identified three types of major systematical
errors. Throughout the paper we use the superscript ``$(o)$'' ($
o=1,2,3,\cdots$) to denote them.

\subsubsection{The systematical error from deterministic bias}
The first set of systematical errors come from errors in determining
the galaxy bias $b_{i}$ through Eq. \ref{eqn:bi} , even if we
neglect the shot noise contribution to it. This bias arises due to a
degeneracy that is among the power spectra $C_{\rm m_bm_b}$, $C_{\rm
m_b\kappa_b}$, $C_{\rm \kappa_b\kappa_b}$ and galaxy bias $b_i$ in
Eq. \ref{eqn:bi}, which causes a slight deviation from its true flux
dependence.  In a companion paper we will address and clarify this
issue in more detail (Yang \& Zhang, in preparation). We will also
show that this systematical error is correctable.

The consequence is  that $\sum_{i} w_{i}^{(n)}b_{i}\neq 0$ in Eq.
\ref{eqn:k}. It causes a systematical error in the auto correlation
of the same redshift bin \ba \label{eqn:dCbb1}\delta C_{\rm
bb}^{(1)}=\left(\sum_{i} w_{i}^{\rm b}b_{i}^{\rm b}\right)^2C_{\rm
m_bm_b}+2\left(\sum_{i} w_{i}^{\rm b}b_{i}^{\rm b}\right)C_{\rm
  m_b\kappa_b}\ .
\ea It also biases the cross correlation measurement between two
different redshift bins, \ba \label{eqn:dCbf1}\delta C_{\rm
bf}^{(1)}=\left(\sum_{i} w_{i}^{\rm f}b_{i}^{\rm f}\right)C_{\rm
m_f\kappa_b}\ . \ea

\begin{figure}
\includegraphics[angle=270,width=84mm]{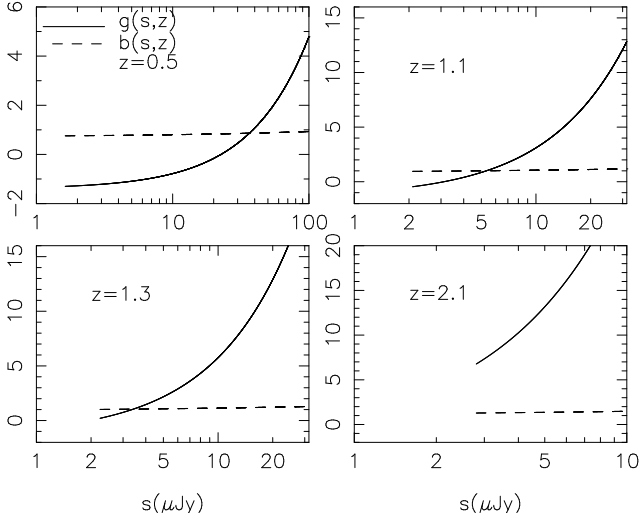}
\caption{The HI galaxy bias $b(s,z)$ and the magnification bias
$g(s,z)=2(\alpha-1)$ as a function of flux $s$ by fixing the
redshift to be the central value of each redshift bin. The plotted
curves are started from the flux limit at the fixed redshift. The
cosmic magnification bias strongly depends on the flux, while the
galaxy bias weakly changes with it. Such difference in flux
dependence ensures us to find a optimal estimator to reconstruct the
weak lensing from the cosmic magnification. } \label{fig:gb_s}
\end{figure}

\subsubsection{The systematical error from stochastic bias}
Our reconstruction has approximated the galaxy bias to be
deterministic, namely $r_{ij}=1$. $r_{ij}$ is the correlation
coefficients between galaxies with different flux. It is known that
galaxy bias exists a stochastic component and hence $r_{ij}<1$,
especially at nonlinear scales
\citep{Wang2007,Swanson2008,Gil-Marin2010}. This does not affect the
determination of the galaxy bias $b_{i}$, since we only use the auto
correlation between the same flux and redshift bin to determine the
bias. However, stochasticity does bias the power spectrum
measurement, since now the condition (Eq. \ref{eqn:wb}) no longer
guarantees a complete removal of the galaxy intrinsic clustering.
The systematical error induced to the auto correlation measurement
is
\begin{eqnarray}
\label{rs} \delta C_{\rm bb}^{(2)}&=&-\left[\sum_{i,j}w_{i}^{\rm
b}w_{j}^{\rm b}b_{i}^{\rm b}b_{j}^{\rm b}\Delta r_{ij}\right]C_{\rm
m_bm_b}.
\end{eqnarray}
Here, $\Delta r_{ij}\equiv 1-r_{ij}$.

Given present poor understanding of galaxy stochasticity, we
demonstrate this bias by adopting a very simple toy model, with
\begin{eqnarray}
\Delta r_{ij}=1-r_{ij}=\left\{
\begin{array}{cc}
0 & (i=j) \\
1\% & ~~(i\neq j)\ .
\end{array}
\right.
\end{eqnarray}
This model is by no way realistic. The particular reason to choose
this toy model is that readers can conveniently scale the resulting
$\delta C^{(2)}$ to their favorite models of galaxy stochasticity by
multiplying a factor $100\Delta r_{ij}(\ell,z)$.

As we will show later, this systematical error could become the dominant error
source. However, measuring the lensing power spectrum between two redshift
bins can avoid this problem. Clearly, stochasticity in galaxy distribution does not bias
such cross power spectrum measurement.
\begin{figure*}
\includegraphics[angle=270,width=160mm]{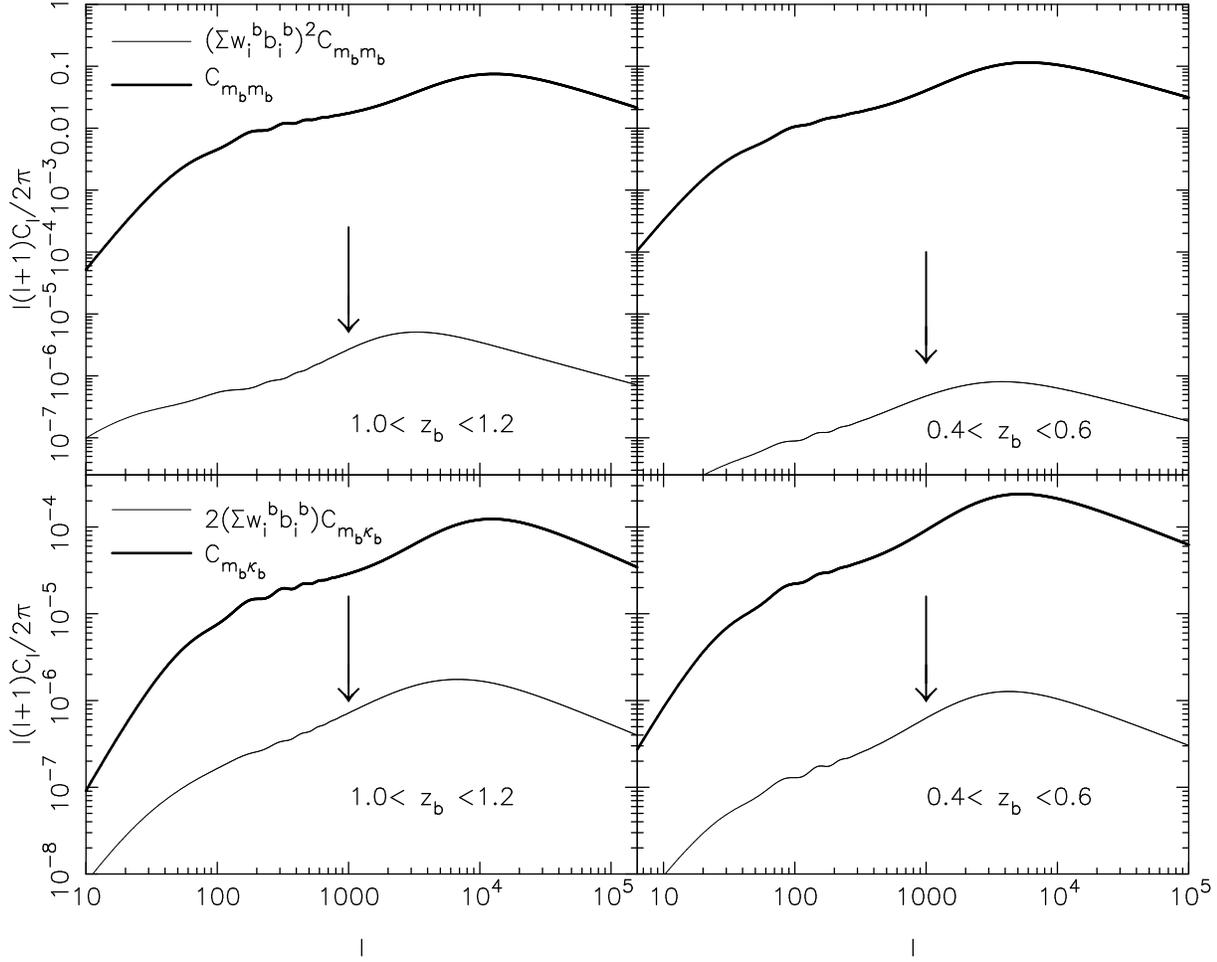}
\caption{The contaminations before and after using the estimator to
reconstruct the weak lensing auto power spectrum. We choose two
redshift bins $0.4<z_{\rm b}<0.6$ and $1.0<z_{\rm b}<1.2$, for
which, the upper panel shows us the suppression of matter auto power
spectra, and the lower panel presents the suppression of
matter-lensing cross power spectra, respectively. Clearly, for the
same redshift bin our reconstruction method can sharply reduce the
two correlations both induced by the intrinsic clustering. The power
spectrum $C_{\rm m_bm_b}$ can be suppressed by an order of
$\sim10^4$ and the power spectrum $C_{\rm m_b\kappa_b}$ can be
reduced by one or more orders of magnitude. The suppression is more
stronger for the lower redshift bin, since the value of
$\sum_{i}w_{i}^{\rm b}b_{i}^{\rm b}$ rises with redshift. Roughly
speaking, this is caused by the increasing error in the final galaxy
bias. With the increasing redshift, the contribution from the weak
lensing power spectrum to the observed power spectrum increases, so
the error in the initial galaxy bias increases and then leads to the
rising error in the final galaxy bias because of the existing
degeneracy in the process of solving galaxy bias. }
\label{fig:suppression auto}
\end{figure*}
\subsubsection{The systematical error from shot noise}
By far we have neglected the influence of galaxy  shot noise in
determining the galaxy bias (Eq.\ref{eqn:bi}). The induced error is
denoted as $\delta b_{j}=b_{j}^{\rm r}-\hat{b_{j}}$. Here,
$b_{j}^{\rm r}$ is the true bias and $\hat{b}_{j}$ is the final
obtained bias $b_{j}^{(n)}$ from the iteration. It is reasonable to
consider the case that shot noise is subdominant to the galaxy
intrinsic clustering. Under this limit,
\begin{equation}
\label{explanation} \left\langle \delta
b_{j}\right\rangle=0,\left\langle (\delta
b_{j})^2\right\rangle\simeq \frac{C_{j}^{\rm shot}}{C_{\rm m_bm_b}}.
\end{equation}
Where $C_{j}^{\rm shot}$ is the shot noise power spectrum of $j$-th
flux bin in the observed power spectrum (Eq. \ref{eqn:bi}). We are
then able to Taylor expand flux weighting $w_{i}$ around
$\hat{b}_{j}$ to estimate the induced bias in it,
\begin{eqnarray}
w_{i}(\hat{b_{j}}+\delta
b_{j})&=&w_{i}(\hat{b_{j}})+\sum_{j}\frac{\partial w_{i}}{\partial
b_{j}}\Big|_{\hat{b_{j}}}\delta b_{j}
\\ \nonumber
&+&\frac{1}{2}\sum_{jk}\frac{\partial^2w_{i}}{\partial b_{j}\partial
b_{k}}\Big|_{\hat{b_{j}}}\delta b_{j}\delta b_{ k}+\cdots
\end{eqnarray}
Where $w_{i}(\hat{b_{j}})$ is the final obtained flux weighting.
After a lengthy but straightforward derivation, we derived the
induced bias in the auto correlation measurement,
\begin{eqnarray}
\label{bs} \delta C_{\rm bb}^{(3)}&=&C_{\rm m_bm_b}\times
\left[\sum_{j}\left\langle (\delta
b_{j}^{\rm b})^2\right\rangle\big(\sum_{i}\frac{\partial w_{i}^{\rm b}}{\partial b_{j}^{\rm b}}b_{i}^{\rm b}\big)^2\right. \\
&&+(\sum_{i} w_{i}^{\rm b}b_{i}^{\rm b})(\sum_{j} \langle (\delta
b_{j}^{\rm b})^2\rangle \sum_{i}\frac{\partial^2w_{i}^{\rm
b}}{\partial
b_{j}^{\rm b}\partial b_{j}^{\rm b}}b_{i}^{\rm b}) \nonumber \\
&&-\left. \sum_{k}\big\langle (\delta b^{\rm
b}_{k})^2\big\rangle\sum_{ij}\frac{\partial w_{i}^{\rm b}}{\partial
b_{k}^{\rm b}}\frac{\partial w_{j}^{\rm b}}{\partial b_{k}^{\rm b}}b_{i}^{\rm b}b_{j}^{\rm b}\Delta r_{ij} \right] \nonumber\\
&+& C_{\rm m_b\kappa_b}\times \left[\sum_{j}\big\langle (\delta
b_{j}^{\rm b})^2\big\rangle\sum_{i}\frac{\partial^2w_{i}^{\rm
b}}{\partial b_{j}^{\rm b}\partial b_{j}^{\rm b}}b_{i}^{\rm
b}\right] \ . \nonumber
\end{eqnarray}
The induced bias in cross correlation measurement is
\begin{eqnarray}
\label{bbf} \delta C^{(3)}_{\rm bf}= \frac{1}{2}C_{\rm
m_f\kappa_b}\times \left[\sum_{j}\big\langle (\delta b_{j}^{\rm
f})^2\big\rangle\sum_{i}\frac{\partial^2w_{i}^{\rm f}}{\partial
b_{j}^{\rm f}\partial b_{j}^{\rm f}}b_{i}^{\rm f}\right] \ .
\end{eqnarray}

\subsection{Other sources of error}
There are other sources of error that we will neglect in the
simplified treatment presented here. First of all, we only deal with
idealized surveys with uniform survey depth without any masks.
Complexities in real surveys will not only impact the estimation of
errors listed in previous sections, but may also induce new sources
of error.  These errors can be investigated with mock catalog
mimicking real observations. We will postpone such investigation
elsewhere.

Another source of error is the determination of $\alpha$ (or equivalently
$g$). For SKA that we will target at,  errors in $\alpha$ are negligible since
the galaxy luminosity function can be determined to high accuracy given
billions of SKA galaxies with spectroscopic redshifts. However, this may not be
the case for other surveys, due to at least two reasons. (1) Some surveys may
not have sufficient galaxies at bright end. Large Poisson fluctuations then
forbidden precision measurement of  $\alpha$ there. (2) $\alpha$ is defined
with respect to the galaxy luminosity function in a given redshift
bin. However, a large fraction of galaxy surveys may only have photometric
redshift measurement. Errors
in redshift, especially the catastrophic photo-z error,
affect the measurement of $\alpha$.

Dust extinction is also a problem for optical surveys, but not for
radio surveys like SKA. Dust extinction also induces fluctuations in
galaxy number density, with a characteristic flux dependence
$\alpha$. This flux dependence differs from the $\alpha-1$
dependence in cosmic magnification and $b_{\rm g}(s)$ dependence in
galaxy bias. The minimal variance estimator can be modified such
that $\sum_{i} w_{i}\alpha_{i}=0$ to eliminate this potential source
of error, when necessary.

In next section, we will quantify statistical and systematical errors listed
in \S \ref{subsec:statisticalerror} \& \ref{subsec:systematicalerror} for the
planned  21cm survey  SKA. Although  the estimation is done under very
simplified conditions, it nevertheless demonstrates that these errors are
likely under control.

\begin{figure}
\includegraphics[angle=270,width=84mm]{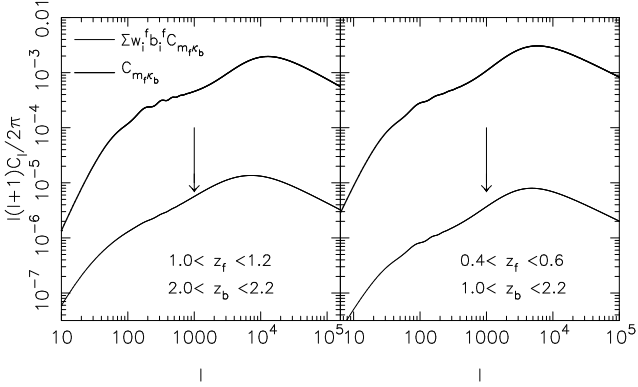}
\caption{The suppression of contamination related with foreground
intrinsic clustering in the reconstructed lensing cross-correlation
power spectrum between foreground and background redshift bins.
Because of the reduced foreground intrinsic clustering, the enormous
correction $C_{\rm m_f\kappa_b}$ to the signal can be suppressed by
a factor $\sim10^2$.} \label{fig:suppression_cross}
\end{figure}

\section{The performance of the minimal variance estimator}
\label{sec:result} We target at SKA to investigate the feasibility
of our proposal. SKA is able to detect billions of galaxies through
their neutral hydrogen 21cm emission. The survey  specifications are
adopted as field of view $\rm FOV=10 \rm deg^2$, total survey period
$\rm t_{\rm all}=5 \rm yr$ and total sky coverage $10^4 \rm deg^2$
\citep{Dewdney2009, Abdalla2010, Faulkner2010}. More details of the
survey are given in the appendix.

Fig. \ref{fig:Cbb} and Fig. \ref{fig:Cbf} demonstrate contaminations
of galaxy intrinsic clustering to the cosmic magnification
measurement from one same redshift bin and one couple of foreground
and background redshift bins. For a typical redshift bin $1.0<z_{\rm
b}<1.2$, the auto matter angular power spectrum $C_{\rm m_bm_b}$ is
larger than the lensing power spectrum  by two or more orders of
magnitude. Fig. \ref{fig:Cbb} also shows $C_{\rm m_b\kappa_b}$ is
comparable to $C_{\rm \kappa_b\kappa_b}$. Since typical bias of 21cm
galaxies is $\sim 1$ (Fig. \ref{fig:gb_s}), this means that the
galaxy intrinsic clustering overwhelms the lensing signal by orders
of magnitude. Similarly, in Fig. \ref{fig:Cbf} the cross power
spectrum $C_{\rm m_f\kappa_b}$ induced by the foreground intrinsic
clustering overwhelms the weak lensing power spectrum $C_{\rm
\kappa_f\kappa_b}$ by one or more orders of magnitude. These big
contaminations related galaxy intrinsic clustering make the weak
lensing measurement difficult from the directly cosmic magnification
measurement, unless for sufficiently bright foreground and
background galaxies at sufficiently high redshifts
\citep{Zhang2006}.

As we explained in earlier sections, the key to extract the lensing
signal from the overwhelming noise is the different dependences of
signal and noise on the galaxy flux. Fig. \ref{fig:gb_s} shows the
lensing signal and the intrinsic clustering indeed have very
different dependences on the galaxy flux. For the 21cm emitting galaxies, the
flux dependence in $g\equiv 2(\alpha-1)$ is much stronger than that in the
galaxy bias. Furthermore,  $g$ changes sign from faint end to bright end. Such
behavior can not be mimicked by bias, which keeps positive.

From such difference in the flux dependences, we expect that our
estimator to significantly  reduce contaminations from the galaxy
intrinsic clustering. As explained earlier, in the galaxy
correlation between the same redshift bin, the intrinsic clustering
induces a systematical error proportional to $C_{\rm m_bm_b}$ (Eq.
\ref{eqn:dCbb1}) and an error proportional to $C_{\rm m_b\kappa_b}$
(Eq. \ref{eqn:dCbb1}). In the ideal case that both the stochasticity
and shot noise in galaxy distribution can be neglected, the
systematical error proportional to $C_{\rm m_bm_b}$ will be
suppressed by a factor $1/(\sum_{i} w_{i}^{\rm b}b_{i}^{\rm b})^2$
(Eq. \ref{eqn:dCbb1}). Fig. \ref{fig:suppression auto} shows that
this suppression factor is of the order $\sim 10^4$ at interesting
scales $10\la \ell\la 10^4$. For the same reason, the systematical
error proportional to $C_{\rm m_b\kappa_b}$ will be suppressed by a
factor $1/\sum_{i} w_{i}^{\rm b}b_{i}^{\rm b}\sim 10^2$ over the
same angular scales. The systematical error induced by foreground
intrinsic clustering in the weak lensing reconstruction between two
redshift bins is proportional to $C_{\rm m_f\kappa_b}$ (Eq.
\ref{eqn:dCbf1}). Our estimator also suppresses it by a factor
 $1/\sum_{i}w_{i}^{\rm f}b_{i}^{\rm f}\sim 10^2$, as shown in Fig.
\ref{fig:suppression_cross}.

\begin{figure*}
\includegraphics[angle=270,width=170mm]{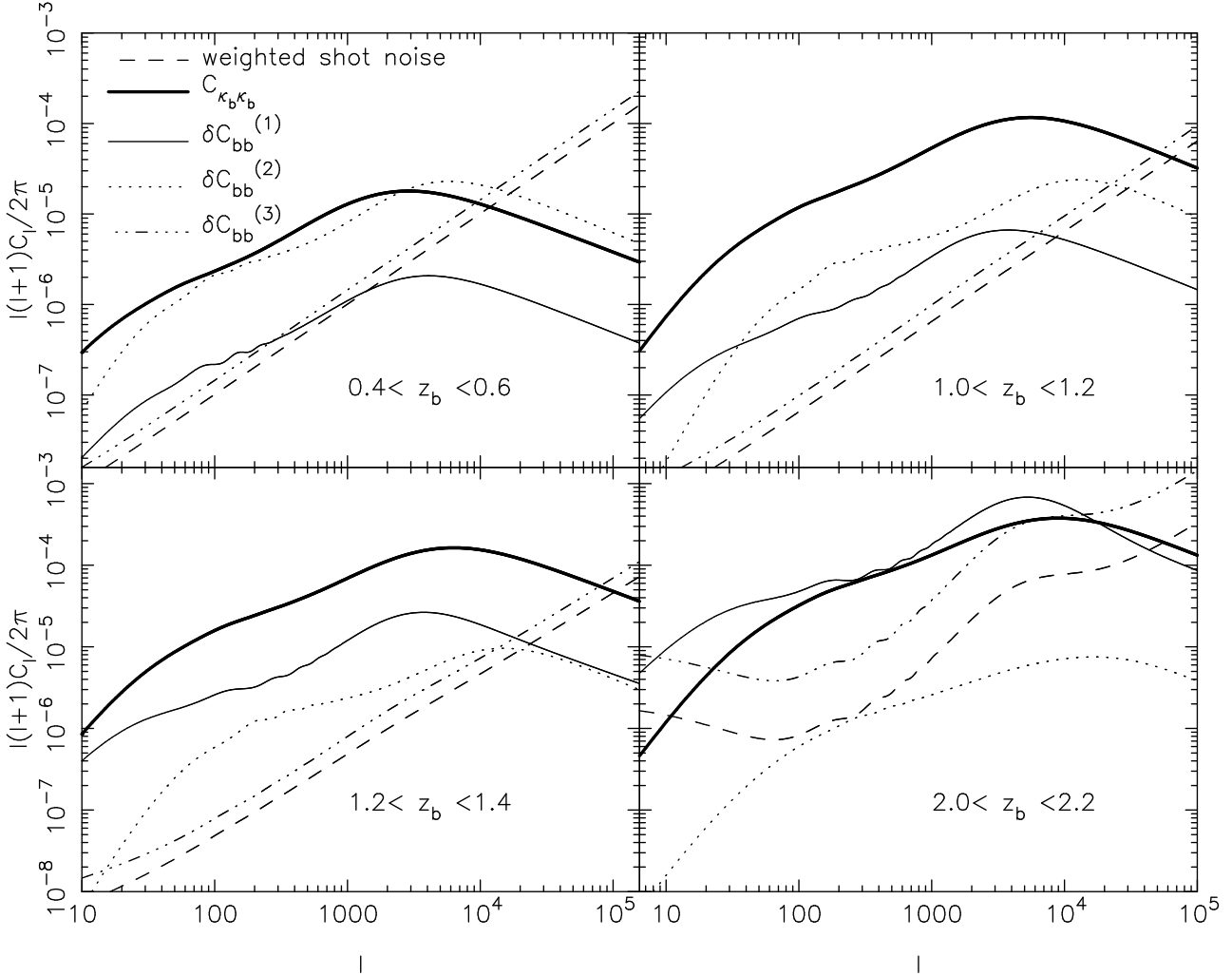}
\caption{Composition of the reconstructed weak lensing power
spectrum for the same redshift bin. We plot the weak lensing  power
spectrum $C_{\rm \kappa_b\kappa_b}$ by bold solid line. The three
types of systematical errors $\delta C_{\rm bb}^{(1)}$ from error in
deterministic galaxy bias, $\delta C_{\rm bb}^{(2)}$ from the
stochastic bias, and $\delta C_{\rm bb}^{(3)}$ from the shot noise
are presented by the solid, dotted and dash-dot-dot-dotted lines,
respectively. While statistical error is plotted by the dashed line.
Here the statistical error is called weighted shot noise only from
the sparse galaxy distribution, since we aim to reconstruct the weak
lensing at each angular pixel with the corresponding cosmic
variance. For the intermediate redshift bins $1.0<z_{\rm b}<1.2$ and
$1.2<z_{\rm b}<1.4$, the signal overwhelms all these errors which
can be controlled to $\sim 10\%$-$20\%$ level.}\label{fig:one}
\end{figure*}

\subsection{Same redshift bins}
 The lensing power spectrum can then be directly measured through the
   reconstructed lensing maps. This can be done for maps of the same redshift
   bin. Fig. \ref{fig:one} compares the residual systematical errors to the lensing
power spectrum signal. Overall, our minimal variance estimator significantly suppresses the
 galaxy intrinsic clustering and makes the weak lensing reconstruction
 feasible.
 For source redshift $z_{\rm b}\sim 1$ (e.g.  $1.0<z_{\rm b}<1.4$), all
investigated  systematical errors and statistical errors are
suppressed to be subdominant to  the signal.  We expect the lensing
power spectrum to be measured with an accuracy of $\sim
10\%$-$20\%$.

However, at lower and higher redshifts, the reconstruction is not
successful.  We observe the systematical error $\delta C^{(1)}_{\rm
bb}$ increases with increasing redshift and overwhelms the lensing
signal at $z_{\rm b}\ga 2$. In the current formalism it is difficult
to explain this behavior straightforwardly. But roughly speaking,
this is caused by worse initial guess on the flux dependence of
galaxy bias coupled with the degeneracy explained earlier. The
initial guess is accurate to the level of $C_{\rm
\kappa_b\kappa_b}/C_{\rm m_bm_b}$. So the associated error increases
with redshift.

This systematical error looks rather frustrating. However in the
companion paper (Yang \& Zhang, in preparation) we will show that
the systematical error $\delta C^{(1)}$ is correctable. Where
 we can separate the degeneracy, and reconstruct a
quantity $y_i^{\rm b}=\sqrt{C_{\rm m_bm_b}}(b_i^{\rm b}+g_i^{\rm
b}C_{\rm m_b\kappa_b}/C_{\rm m_bm_b})$ through a direct
multi-parameter fitting against the measured power spectra, which
perfectly mimics the flux dependence of galaxy bias $b_i^{\rm b}$,
because the power spectrum $C_{\rm m_bm_b}$ is independent with the
flux and the correction term $C_{\rm m_b\kappa_b}/C_{\rm m_bm_b}$ is
small especially for high redshift. Although it is bad to find that
the final convergence depends on the initial guess of galaxy bias,
$y_i^{\rm b}$ as a guess of galaxy bias does  reduce the
systematical error $\delta C^{(1)}$ and then works far better than
the obtained galaxy bias in this paper.

The systematical error $\delta C^{(3)}$ arising from shot noise
becomes non-negligible at $z_{\rm b}\sim 2$, due to sharply
decreasing galaxy density and increasing shot noise at these
redshifts (see Fig. \ref{fig:dn_dz} ). We find that this error is
always subdominant to either $\delta C^{(1)}_{\rm bb}$ or $\delta
C^{(2)}_{\rm bb}$.   Interestingly, both $\delta C^{(3)}_{\rm bb}$
and the
  weighted shot
noise $\Delta C_{\rm bb}$ (statistical error, Eq.
\ref{eqn:statistical}) have similar shapes at all redshifts.
Furthermore, both roughly scale as $l^0$ at low redshifts. These are
not coincidences. (1) The term $\propto C_{\rm m_bm_b}$ is dominant
in $\delta C^{(3)}_{\rm bb}$. This term is also $\propto \langle
(\delta b^{\rm b}_{j})^2\rangle\propto 1/C_{\rm
  m_bm_b}/\bar{n}_{j}$. So both $\delta C^{(3)}_{\rm bb}$ and $\Delta C_{\rm
  bb}$ are the sums of $1/\bar{n}_{j}$ weighted in different
ways and hence have similar shapes.  (2) The galaxy bias in our
fiducial model is scale independent. This results in scale
independent weighting function $w_{j}$, as long the bias can be
determined to high accuracy. For these reasons, $\delta C^{(3)}_{\rm
bb}, \Delta C_{\rm bb}\propto l^0$. This is the case at low
redshift. (3) However, the accuracy in determining galaxy bias is
significantly degraded by contamination proportional to
$C_{\kappa_{\rm b}\kappa_{\rm b}}/C_{\rm
  m_bm_b}$, which is scale dependent and increases with redshift. This is the
reason both  $\delta C^{(3)}_{\rm bb}$ and $\Delta C_{\rm
  bb}$ show complicated angular dependence at $z_{\rm b}\simeq 2$
(Fig. \ref{fig:one}).

At low redshift (e.g. $0.4<z_{\rm b}<0.6$), the systematical error
$\delta C^{(2)}_{\rm bb}$ arising from galaxy bias stochasticity
becomes dominant or even exceeds the lensing signal. This is what
expected. The galaxy intrinsic clustering is stronger at lower
redshift. This amplifies the impact of galaxy stochasticity. Weaker
lensing signal at lower redshift further amplifies its relative
impact.  As expected, Fig. \ref{fig:one} shows that $\delta
C^{(2)}_{\rm bb}/C_{\kappa_{\rm
    b}\kappa_{\rm b}}$ decreases with increasing redshift and becomes negligible
at $z_{\rm b}\sim 2$.    Since $\delta C^{(2)}_{\rm bb} \propto
\Delta r_{ij}$, the importance of $\delta C^{(2)}_{\rm bb}$ is
sensitive to the true nature of galaxy stochasticity. Its value
should be multiplied by a factor $100 \Delta r_{ij}$ for fiducial
value of $\Delta r_{ij}\neq 1\%$.  We hence conclude that galaxy
stochasticity is likely the dominant source of error in weak lensing
reconstruction through cosmic magnification.

\begin{figure*}
\includegraphics[angle=270,width=170mm]{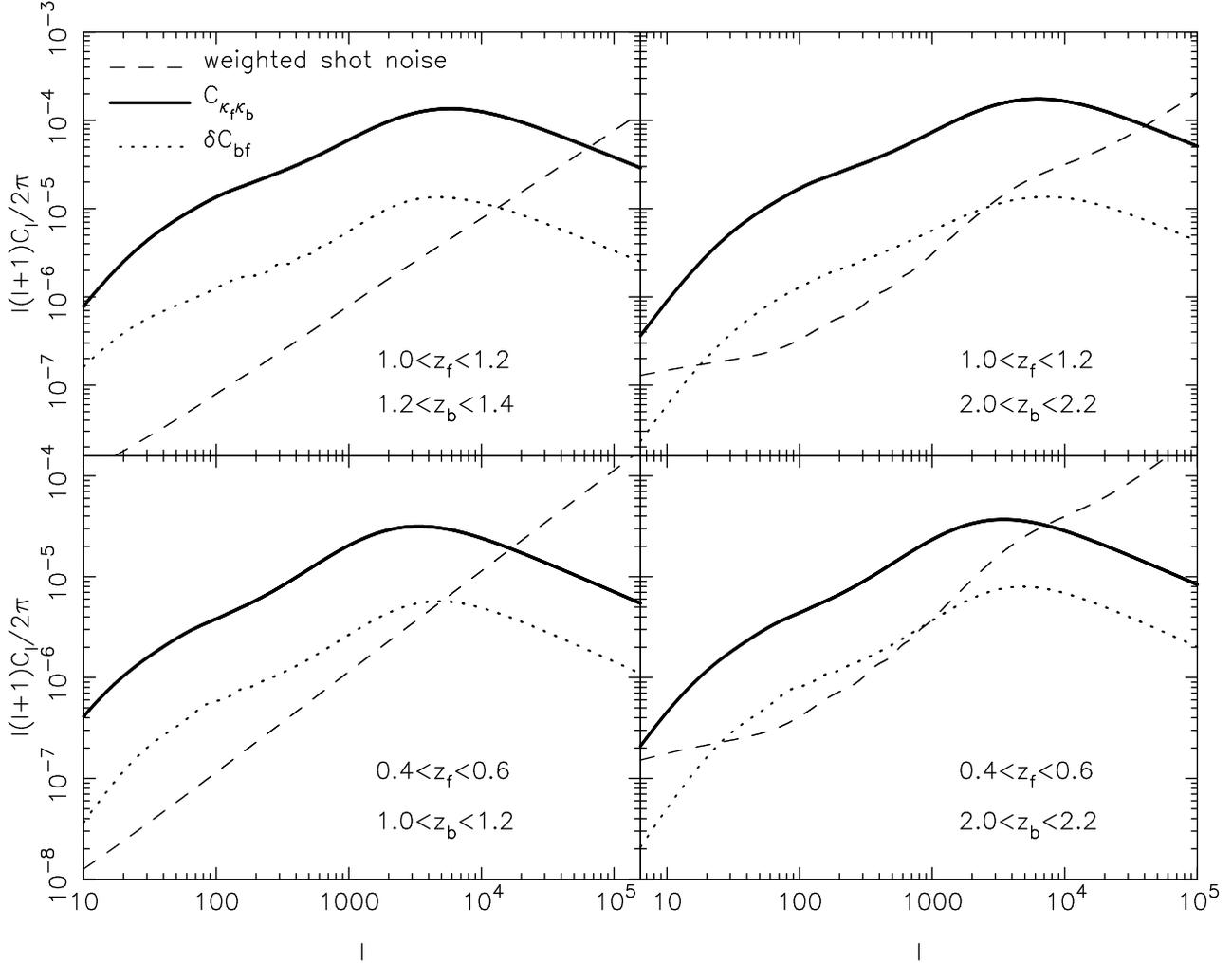}
\caption{The weak lensing signal, the systematical error and the
statistical error in the weak lensing reconstruction from the
foreground and background redshift bins. The solid line is the
cross-correlation power spectrum of cosmic convergence. The dotted
line represents the systematical error combing three types $\delta
C_{\rm bf}=\delta C_{\rm bf}^{(1)}+\delta C_{\rm bf}^{(2)}+\delta
C_{\rm bf}^{(3)}$ and the dashed line corresponds to the weighted
shot noise. In $\delta C_{\rm bf}$, $\delta C_{\rm bf}^{(1)}$ from
the deterministic galaxy bias is dominant and the systematical error
from the stochastic bias can be avoided in such cross lensing power
spectrum, namely $\delta C_{\rm bf}^{(2)}=0$. For every couple of
foreground and background redshift bins, the cross reconstructed
weak lensing signal dominates at scale range $10\la \ell\la 10^4$
and it can be measured to reach $\sim 10\%$-$20\%$ accuracy.}
\label{fig:two}
\end{figure*}

\subsection{Different redshift bins}
 Fortunately this stochasticity issue can be safely overcome in the lensing
power spectrum measurement through lensing maps reconstructed in two different
redshift bins (foreground and background bins). The results are shown in Fig.
\ref{fig:two}. In this case, the systematical error is dominated by $\delta
C^{(1)}_{\rm bf}$. The stochasticity does not induce systematical error so
that $\delta C^{(2)}_{\rm bf}=0$.

Overall, the lensing power spectrum measurement through cross
correlating reconstructed maps of different redshift bins is more
robust than the one based on the same redshift bin. The
reconstruction accuracy can be controlled to $10\%$-$20\%$ over a
wide range of foreground and background redshifts.

At last, for a consistency test, we check whether our results depend on the
division of the  flux bin. As expected,  various systematical errors and
statistical error change little with respect to different choices of flux
bins, as long as these bins are sufficiently fine.

\subsection{Uncertainties in the forecast}
 There are a number of uncertainties in the forecast, besides the ones
  discussed in previous sections. (1) In the fiducial
  galaxy intrinsic clustering model, we have
  neglected the scale dependence in galaxy bias.  (2) We have neglected cosmic
  variance in the lensing signal so the fiducial power spectrum is the
  ensemble average. But we do not expect it can
significantly impact our result, since the cosmic variance at most
relevant scale is small due to $l\Delta lf_{\rm sky}\gg1$ with a
given large sky coverage of SKA. (3) The toy model of galaxy
stochasticity is too
  simplified. In reality, the cross correlation coefficient $r$ should be a
  function of redshift, angular scale and galaxy type and flux.

For these reasons, the numerical results presented in this paper
should only be trusted as rough estimation. Robust evaluation of the
weak lensing reconstruction performance through cosmic magnification
requires much more comprehensive investigation. Nevertheless, the
concept study shown in this paper demonstrate that weak lensing
reconstruction through cosmic magnification is indeed promising.

\section{Conclusions and discussions}
\label{sec:conclusion} We propose a minimal variance estimator to
reconstruct the weak lensing  convergence $\kappa$ field through the
cosmic magnification effect in the observed  galaxy number density
distribution. This estimator separates the galaxy intrinsic
clustering from the lensing signal due to their distinctive
dependences on the galaxy flux. Using SKA as an example, we
demonstrate the applicability of our method, under highly simplified
conditions. It is indeed able to significantly reduce systematical
errors. We have identified and quantified residual systematical
errors and found them in general under control.  Extensive efforts
shall be made to test our reconstruction method more robustly and to
improve this method.

Comparing to previous works, our method has
several key features/advantages.
\bi
\item  Unlike weak lensing reconstruction
through cosmic  shear,  it does
not involve galaxy shape measurement and reconstruction and hence avoids all
potential problems associated with galaxy shapes.  Hence the reconstructed
lensing maps provide useful independent check against cosmic shear
measurement.
\item Unlike existing cosmic
magnification measurements which actually measure the galaxy-lensing
cross correlation, our estimator allows directly reconstruction of
the weak lensing $\kappa$ field. From the reconstructed $\kappa$, we
are able to directly measure the lensing power spectra of the same
source redshift bin and between two redshift bins. These statistics
do not involve galaxy bias, making them more robust cosmological
probes. The usual lensing tomography is also directly applicable.
\item Unlike our previous works
\citep{Zhang2005,Zhang2006}, the new method does not require priors
on the galaxy bias, especially its flux dependence. Our methods is
able to simultaneously measure the galaxy bias (scaled with a flux
independent factor) and the lensing signal. Hence we do not adopt
any priors on the galaxy bias (other than that it is deterministic)
and treat the galaxy bias as a free function of scale and flux. The
price to pay is degradation in constraining galaxy bias and in
lensing reconstruction. Adding priors on the galaxy bias can further
improve the reconstruction precision, although the reconstruction
accuracy will be affected by uncertainties/biases in the galaxy bias
prior. For this reason, we do not attempt to add priors on galaxy
bias in the reconstruction.
\item Our method is complementary to a recent proposal by \cite{Heavens2011},
which proposes a nulling technique to reduce the galaxy
  intrinsic clustering by proper weighting in redshift.  Comparing to this
  method, our method only utilizes extra information encoded in the flux
  dependence  to reduce/remove the galaxy intrinsic clustering. It
  keeps the cosmological information encoded in the redshift dependence
  disentangled from the process of removing the intrinsic clustering.
\ei

The proposed approach is not the only way for weak lensing
reconstruction through cosmic magnification. The current paper
focuses on direct reconstruction of the lensing convergence $\kappa$
map. In a companion paper, we will focus on direct  reconstruction
of the lensing power spectrum (Yang \& Zhang, in preparation). We
will show that combining two-point correlation measurements between
all flux bins, the lensing power spectrum can be reconstructed free
of assumptions on the galaxy intrinsic clustering.  We will see that
this approach is more straightforward, more consistent and easier to
carry out. However the method presented in this paper does have
advantages. Since it reconstructs the lensing $\kappa$ map, higher
order lensing statistics such as the bispectrum can be measured
straightforwardly. Furthermore, the reconstructed $\kappa$ map can
be straightforwardly correlated with other tracers of the large
scale structure. For example, it can be correlated with the  lensing
map reconstructed from CMB lensing  \citep{Seljak1999, Hu2002} or
21cm lensing. Furthermore, through this approach we can have better
understanding on the origin of various systematical errors, which
can be entangled in the alternative approach.

 Our reconstruction method is versatile to include other components of
  fluctuation in the galaxy number density. The extinction induced fluctuation
  discussed earlier is one. High order corrections to the cosmic magnification
  is another.  Taylor expanding Eq. \ref{eqn:n} to the second order, we obtain \ba
\delta_g^L&=&\delta_g+2(\alpha-1)\kappa\\
&&+2(\alpha-1)\left[(\kappa\delta_g-\langle\kappa\delta_g\rangle)+\frac{1}{2}(\gamma^2-\left\langle\gamma^2\right\rangle)\right]
\nonumber \\
&&+(1-5\alpha+2g_2)\left[\kappa^2-\left\langle\kappa^2\right\rangle\right]+O\left[\kappa^3,\gamma^3,\cdots\right]\ . \nonumber
\ea
Here $g_2\equiv (s^2/n)d^2n/ds^2$ is related with the
second derivative of luminosity function.

The above result shows that the $\kappa$ reconstruction through cosmic
magnification is biased by terms proportional to $\kappa \delta_g$ and
$\gamma^2$ (second line in the above equation).  Similar biases also exist in cosmic shear measurement. We
recognize $\kappa \delta_g$ as the source-lens coupling. The $\gamma^2$ term
is analogous to the $\kappa\gamma$ term caused by reduced shear
$\gamma/(1-\kappa)$.  Precision lensing cosmology has to model these
corrections appropriately.

The high order corrections $\propto 1-5\alpha+2g_2$ can in principle
be separated due to  its unique flux dependence. However, it is
unclear whether the reconstruction is doable, even for a survey as
advanced as SKA.

\section*{acknowledgments}
This work is supported by the one-hundred talents program of the
Chinese academy of science, the national science foundation of China
(grant No. 10821302, 10973027, 11025316 \& 10973018),  the CAS/SAFEA
International Partnership Program for  Creative Research Teams and
National Basic Research Program of China (973 Program) under grant
No.2009CB24901.

\appendix
\section[]{SKA survey}
\label{sec:SKA} SKA is a future radio survey with aiming to
construct the world's largest radio telescope. Through the neutral
hydrogen emitting 21 cm hyperfine transition line, it can observe
large sample of 21 cm galaxies. Even at high-redshift, the observed
21 cm galaxies is extremely excess than the QSOs or LBGs that is
usually used as background object \citep{Scranton2005,
Hildebrandt2009}, so it is easily to overcome the shot noise in the
measurement of cosmic magnification effect, of which precise
measurement is sensitive to the shot noise. In addition, compared
with photometric survey, the spectroscopic survey can determine the
redshift more precise by using the redshifted wavelength of 21 cm
line ($\lambda=\lambda_0(1+z)$). Furthermore, radio survey is free
of galactic dust extinction, which is correlated with foreground
galaxies and then induces a correction to the galaxy-galaxy cross
correlation. As a consequence, it is expected a good performance of
our proposed method to measure the cosmic magnification for SKA. In
this paper, we adopt the survey specifications as follows: the
telescope field-of-view (FOV) is $\rm 10 deg^2$ without evolution,
the overall survey time is $\rm 5 yr$ and the sky coverage is $\rm
10000 deg^2$.
\subsection{HI mass limit}
By assuming a flat profile for the emission line with frequency, we
obtain the following equation linking $M_{\rm HI}$ with the observed
flux density $s$ by atomic Physics (detail in \cite{Abdalla2005}),
\begin{equation}
M_{\rm HI}=\frac{16\pi}{3}\frac{m_H}{A_{21}hc}\chi^2(z)sV(z)(1+z)
\label{eqn:M_HI}\ .
\end{equation}
Where $\chi(z)$ is the comoving angular distance, $A_{21}$ is
spontaneous transition rate, $m_H$ is the atomic mass of hydrogen,
$h$ is Planck's constant and $V(z)$ is the line-of-sight velocity
spread. We assume a scaling with redshift for the typical rest-frame
velocity width of the line $V(z)=V_0/\sqrt{1+z}$, where $\rm
V_0=300kms^{-1}$, and ignore the effects of inclination. The r.m.s.
sensitivity  for a dual-polarization radio receiver at system
temperature $\rm T_{sys}$ for an integration of duration $t$ on a
telescope of effective collecting area $\rm A_{eff}$ is given by
\begin{equation}
s_{\rm rms}=\frac{\sqrt{2}T_{\rm sys}}{\eta_cA_{\rm
eff}}\frac{k_B}{\sqrt{\Delta\nu t}} \label{s_{rms}}\ .
\end{equation}
Here, the correlation efficiency $\eta_c$ is adopted as
$\eta_c=0.9$, the duration time $t=40h$ for each area of the sky and
the width of the HI emission line determines the relevant frequency
bandwidth $\Delta\nu$, which is related to a line-of-sight velocity
width $V(z)$ at redshift $z$
\begin{equation}
\Delta\nu=\frac{\nu_0}{1+z}\frac{V(z)}{c}\ .
\end{equation}
The flux detection limit $s_{\rm lim}$ for galaxy is defined by the
threshold parameter $n_\sigma=\rm s_{lim}/s_{rms}$. Here, we apply
$n_\sigma=5$. From Eq . (\ref{eqn:M_HI}), we can obtain the HI mass
limit.

\begin{figure}
\includegraphics[angle=270,width=84mm]{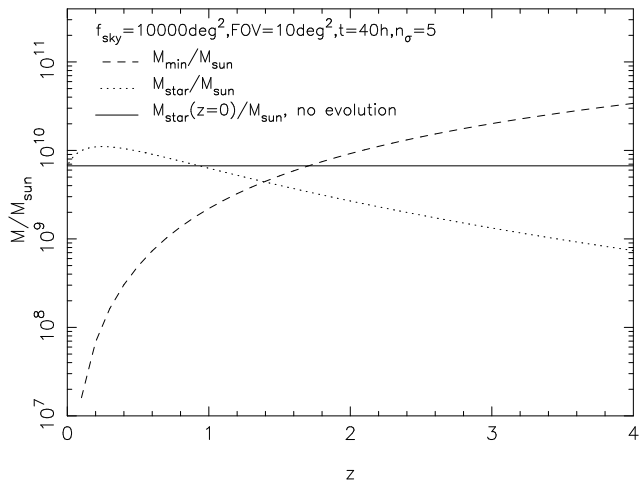}
\caption{ The redshift evolutions of HI mass limit (dashed line) and
the characteristic mass in model C (dotted line). The characteristic
mass in non-evolution model is plotted by the solid line. }
\label{fig:m_evolution}
\end{figure}

\subsection{HI mass function}
we assume that the HI mass function at all redshifts is described by
a Schechter function,
\begin{equation}
\phi(M_{\rm HI},z)dM_{\rm HI}=\phi^\ast\left(\frac{M_{\rm
HI}}{M^\ast}\right)^\beta\exp\left(-\frac{M_{\rm
HI}}{M^\ast}\right)\frac{dM_{\rm HI}}{M^\ast}\ .
\end{equation}
Here, the parameter $\beta$ is low-mass slope, $M^\ast$ is
characteristic mass and $\phi^\ast$ is normalization. The HIPASS
survey reported the results: $\beta=-1.3$, $ M^\ast(z=0)=3.47\rm
h^{-2}10^9M_\odot$ and $\phi^\ast=0.0204\rm h^3Mpc^{-3}$ (Zwaan et
al. 2003). There is little solid measurement of $\phi_\ast(z)$ and
$M_\ast(z)$ other than at local universe. But for this form of  mass
function, there exists a tight relation between $\Omega_{\rm HI}$,
the present day critical density $\rho_c(z=0)$ and
$\phi^\ast(z)M^\ast(z)$,
\begin{equation}
\Omega_{\rm HI}h=\Gamma(\beta +2)\phi^\ast M^\ast(z)/\rho_c(z=0)\ .
\end{equation}
Where $\Gamma$ is the Gamma function. Observation of damped
Lyman-alpha (DLA) system and Lyman-alpha limit system can be used to
measure $\Omega_{\rm HI}$ \citep{Peroux2001, Peroux2003,
Peroux2004}. We use the following functional form produced by
fitting to DLA data points used in paper \citep{Abdalla2010}:
\begin{equation}
\Omega_{\rm HI}=N\left[1.813-1.473(1+z)^{-2.31}\right]\ .
\end{equation}
The normalization constant $N$ is fixed by the value of
$\phi^\ast(z=0)M^\ast(z=0)$.

\begin{figure}
\includegraphics[angle=270,width=84mm]{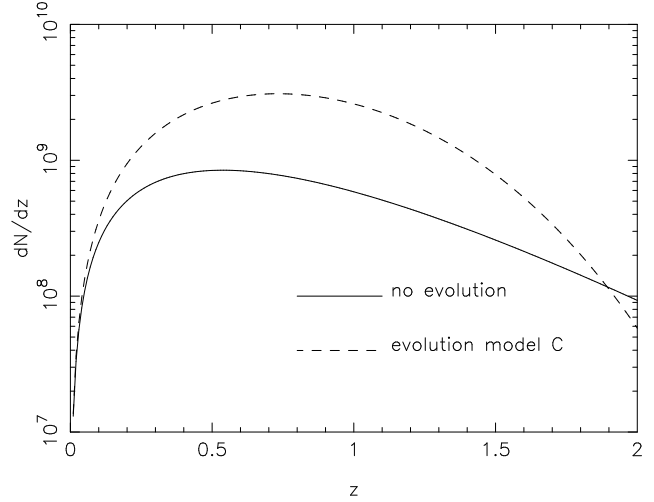}
\caption{The redshift distributions of HI galaxies for no-evolution
model and evolution model C. } \label{fig:dn_dz}
\end{figure}

Here, we adopt model C in papers \citep{Abdalla2005, Abdalla2010} as
 evolution model of HI mass function. In this model, $\phi^\ast$ scales
with $z$ by using the DLA results. The break in the mass function
$M^\ast$ is controlled by the cosmic star formation rate and given
by
\begin{equation}
\frac{M^\ast(z)}{M^\ast(0)}=\left(\frac{\Omega_{\rm
star}(0)/\Omega_{\rm HI}(0)+2}{\Omega_{\rm star}(z)/\Omega_{\rm
HI}(z)+2}\right)D^3(z)\ .
\end{equation}
Where it is assumed  that $\Omega_{\rm HI}(z)=\Omega_{\rm H_2}(z)$
and $D(z)$ is the growth factor. The fractional density in stars
$\Omega_{\rm star}(z)=\rho_{\rm star}(z)/\rho_c(z=0)$ is deduced by
using the fits to the cosmic star-formation history in
\cite{Choudhury2002},
\begin{equation}
\Omega_{\rm star}(z)=\frac{1}{\rho_c(z=0)}\int^{\infty}_z{\frac{{\rm
SFR}(z')}{H(z')(1+z')}dz'}\ ,
\end{equation}
and
\begin{equation}
{\rm
SFR}(z)=\frac{0.13}{1+6\exp(-2.2z)}\left(\frac{h}{0.5}\right)\left(\frac{1}{2.5}\right)\frac{\chi_{\rm
fid}(z)}{\chi(z)}\ .
\end{equation}
Here, $\chi(z)$ and $\chi_{\rm fid}(z)$ are the co-moving distances
at redshift $z$ for the adopt cosmology model and a fiducial
cosmology model with $\Omega_M=1$ and $\Omega_\Lambda=0$,
respectively. The units of SFR are $\rm h^2M_\odot yr^{-1}Mpc^{-3}$.
Fig . \ref{fig:m_evolution} shows us the evolutions of HI mass limit
and characteristic mass in evolution model C. The no-evolution
characteristic mass is also plotted for comparison. In Fig .
\ref{fig:dn_dz}, we plot the redshift distributions of HI galaxy for
the evolution model C and the no-evolution model.

\subsection{The slope of the HI galaxy number count}
In Eq . \ref{a}, $n$ is the mass function of HI galaxies. We can
derive the parameter $\alpha$ as a function of flux $s$ and redshift
$z$, which is
\begin{equation}
\alpha(M_{\rm HI},z)=-\beta-1+\frac{M_{\rm
HI}}{M^\ast(z)}=\alpha(s,z)\ .
\end{equation}
The relationship between flux $s$ and $M_{\rm HI}$ is given by Eq .
\ref{eqn:M_HI}.

\subsection{The galaxy bias $b_g$ for HI galaxy}
In our forecast, we adopt a deterministic bias $b$ by assuming the
cross correlation coefficient $r=1$, and hence we estimate the error
induced by this assumption.

The large scale bias of HI galaxies estimated by method predicted in
paper \citep{Jing1998}.
\begin{eqnarray}
b(M_{\rm DM},z)=\bigg(\frac{0.5}{\nu^4}+1\bigg)^{(0.06-0.02n_{\rm
eff})}\left(1+\frac{\nu^2-1}{\delta_c}\right)\ .
\end{eqnarray}
Where $\nu=\delta_c(z)/\sigma(M_{\rm DM})$. $\sigma(M_{\rm DM})$ is
the linearly evolved $\rm rms$ density fluctuation with a top-hat
window function and $\delta_c(z)=1.686/D(z)$. The effective index is
defined as the slope of $P(k)$ at the halo mass $M_{\rm DM}$,
\begin{equation}
n_{\rm eff}=\frac{dlnP(k)}{dlnk}\left|_{k=\frac{2\pi}{R}}\right . ;
R=\left(\frac{3M_{\rm DM}}{4\pi\rho_c(0)}\right)^{\frac{1}{3}}\ .
\end{equation}
HI galaxies are selected by their neutral hydrogen mass $M_{\rm
HI}$. To calculate the associated biases of these galaxies, we need
to convert $M_{\rm HI}$ to the galaxy total mass $M_{\rm DM}$. Since
most baryons and dark matter are not in galaxies while most neutral
hydrogen atoms are in galaxies, we expect that $f_{\rm
HI}\gg\Omega_{\rm HI}/\Omega_m$. Here, we choose $f_{\rm HI}=0.1$.
Although there are some problems to model the HI galaxy bias by this
way, it is emphasized that our reconstruction method is not limited
to the specific  value of galaxy bias adopted. As long as the galaxy
bias has different flux dependence with the magnification bias.
Their different flux dependences are presented in Fig .
\ref{fig:gb_s}.

\end{document}